\documentclass[useAMS,usenatbib]{mn2e}
\usepackage{graphicx}
\usepackage{color}
\usepackage{soul} %%%\st{Hellow world}
\usepackage{url}
\usepackage{txfonts}
\usepackage{longtable}

\definecolor{myred}{rgb}{0.7,0.0,0.2}

\definecolor{myblue}{rgb}{0.0,0.2,0.7}

\definecolor{mygreen}{rgb}{0.2,0.7,0.0}

\title[Chemical abundance analysis of symbiotic giants. I. RW Hya and SY Mus]{Chemical abundance analysis of symbiotic giants. \\ I. RW Hya and SY Mus}
\author[J. Miko{\l}ajewska et al.]{Joanna
Miko{\l}ajewska$^{1}$\thanks{E-mail: mikolaj@camk.edu.pl}, Cezary Ga{\l}an$^{1}$, Kenneth H.  Hinkle$^{2}$, Mariusz Gromadzki$^{3}$,
\newauthor % starts a new line
and Miros{\l}aw R. Schmidt$^{4}$
\\
$^{1}$N. Copernicus Astronomical Center, Bartycka 18, PL-00-716 Warsaw, Poland\\
$^{2}$National Optical Astronomy Observatory, P.O. Box 26732, Tucson, AZ 85726, USA\\
$^{3}$Departamento de F{\'i}sica y Astronom{\'i}a, Universidad de Valpara{\'i}so, Av. Gran Breta\~na 1111,
Playa Ancha, Casilla 5030, Chile\\
$^{4}$N. Copernicus Astronomical Center, Rabia\'nska 8, PL-87-100 Toru\'n, Poland}

\begin{document}

\date{Accepted 2014 March 11 Received 2014 February 6}

\pagerange{\pageref{firstpage}--\pageref{lastpage}} \pubyear{2014}

\maketitle

\label{firstpage}

\begin{abstract}
The study of symbiotic systems is of considerable importance in our
understanding of binary system stellar evolution in systems where mass loss
or transfer takes place.  Elemental abundances are of special significance
since they can be used to track mass exchange.  However, there are few
symbiotic giants for which the abundances are fairly well determined.  Here
we present for the first time a detailed analysis of the chemical
composition for the giants in the RW Hya and SY Mus systems.  The analysis
is based on high resolution (R $\sim$ 50000), high S/N, near-IR spectra. 
Spectrum synthesis employing standard LTE analysis and atmosphere models was
used to obtain photospheric abundances of CNO and elements around the iron
peak (Sc, Ti, Fe, and Ni).  Our analysis reveals a significantly sub-solar
metallicity, [Fe/H]$\sim$\mbox{-0.75}, for the RW\,Hya giant confirming its
membership in the Galactic halo population and a near-solar metallicity for
the SY\,Mus giant.  The very low $^{12}$C/$^{13}$C isotopic ratios,
$\sim$6--10, derived for both objects indicate that the giants have
experienced the first dredge-up.  \end{abstract}


%==================================================================

\section{Introduction}

Binary systems are an invaluable source of knowledge of the physical
parameters of the stars. During some stage of stellar evolution
most binary stellar systems undergo interactions between the
components.  The interactions turn on and off at various evolutionary
stages, depending on the separation of components. Among the most
interesting examples of such systems are the symbiotic systems.
These are still often called symbiotic stars because their binary
nature which we now consider obvious was controversial only thirty
years ago.

Symbiotic systems are long-period interacting binaries in which an
evolved giant transfers material to a hot and luminous companion
surrounded by an ionized nebula.  The hot component of the vast
majority of symbiotic systems is a white dwarf (WD) although in two
cases, V2116\,Oph \citep{Dav1977} and V934\,Her \citep{Gar1983}, a
neutron star has been found. There are two distinct classes of
symbiotic binaries. The S-type (stellar) have normal red giants and
orbital periods in the range $\sim$1--15 years. The D-type (dusty)
have Mira primaries usually surrounded by a warm dust shell and
orbital periods of typically decades to centuries.  Mass exchange
is the property that defines the symbiotic class.  Even in the
D-type systems the components are still close enough to allow the
WDs to accrete material from the giant's stellar wind.  Symbiotic
systems are the interacting binaries with the longest orbital
periods.  Their study is essential to understand the evolution and
interaction of detached and semi-detached binary stars.

The rich and luminous circumstellar environment surrounding the
interacting symbiotic stellar members results from the presence of
both an evolved giant with a heavy mass loss and a hot companion
copious in ionizing photons often producing its own wind.  The cool
giant and the hot dwarf produce strongly different environments
such as ionized and neutral regions, dust forming region(s),
accretion/excretion disks, interacting winds, bipolar outflows, and
jets.  Such a complex multi-component structure makes symbiotic
stars a very attractive laboratory to study many aspects of stellar
evolution in binary systems.  Firming links between symbiotic systems
and related objects are essential to the understanding, for instance,
of the role of interacting binaries in the formation of stellar
jets, planetary nebulae, novae, supersoft X-ray sources (SSXS), and supernovae type Ia (SN Ia).  Many of these
issues concerning the late stages of stellar evolution are presently
poorly understood but have important implications on our understanding
of the stellar population and chemical evolution of galaxies as
well as the extragalactic distance scale.

Chemical composition is one of the major parameters along with
initial mass that determine stellar evolution.  Chemical composition
has a strong influence on many important astrophysical processes.
In the case of symbiotic stars it is generally believed that the
symbiotic appearance and activity is triggered by high mass-loss
rate of the giant \citep[see e.g.][and references therein]{Mik2003}
possibly due to its enhanced metallicity \citep{Jor2003}.

The apparent deficit of extrinsic C or S stars, i.e. cool components
polluted by the s-process and C-rich material from the former TP-AGB
companion, is among the most interesting problems raised by the
S-type symbiotic binaries.  There are other red giant -- white dwarf
binary star families with abundance peculiarities, for instance the
barium stars and Tc-poor S stars.  These families have orbital
parameters similar to those of symbiotic systems but do not exhibit
symbiotic activity.  The hot component of most symbiotic systems
is a white dwarf and, at least in some systems, the white dwarf
mass is higher than 0.5~$M_\odot$ \citep{Mik2003}.  White dwarfs
of this type have gone through the TP-AGB phase.  Moreover, the
orbital periods of the S-symbiotic systems are generally less than
$\sim$1000 days with circular orbits.  Interaction of these systems
is via Roche-lobe overflow.  \citet{Jor2003} suggested that either
the former AGB star did not go through the TP-AGB or its high
metallicity hindered the efficiency of the s-process and the mass
transfer during the TP-AGB.  The first possibility could apply to
a small number of red symbiotic systems with low mass ($M_{\rm WD}$
$\sim$0.4 $M_\odot$) companions such as AX\,Per and CI\,Cyg. However,
in order to account for the absence of the barium syndrome in systems
like AR\,Pav ($M_{\rm WD}$ $\sim$1 $M_\odot$) or FN Sgr ($M_{\rm
WD}$ $\sim$0.7 $M_\odot$) the second possibility must be invoked.

\begin{table}
 \centering
  \caption{Parameters of RW\,Hya and SY\,Mus}\label{T1}
  \begin{tabular}{@{}lccl@{}}
  \hline
 Parameter        & RW\,Hya         & SY\,Mus         & Reference\\
 \hline
 $P_{\rmn{orb}}$  & 370.2 days      & 625 days        & \citet{KeMi1995}\\
                  &                 &                 & \citet{Dum1999} \\
 Sp Type          & M2              & M5              & \citet{Bel2000} \\
 $M_{\rmn{cool}}$ & 1.6$\pm$0.3     & 1.30$\pm$0.25   & \citet{Rut2007} \\
 $M_{\rmn{hot}}$  & 0.48$\pm$0.06   & 0.43$\pm$0.05   & \citet{Rut2007} \\
 $R_{\rmn{cool}}$ & 122\,$R_{\sun}$ & 135\,$R_{\sun}$ & \citet{Rut2007} \\
\hline
\end{tabular}
\end{table}

There are contradictory arguments provided by the
photospheric abundance analysis of symbiotic giants.  In particular,
the so-called yellow symbiotic systems, AG Dra, BD-21 3873, and
He2-467, all contain K-type giants that are metal poor and s-process
over abundant (\citealt[][1997;]{Smi1996} \citealt{Per1998}).  These
systems seem to belong to the Galactic halo and are probably
low-metallicity relatives of Ba stars.  On the other hand, HD\,330036,
AS\,201, and StHA\,190, members of another small subclass of symbiotic
stars with G-type giants and warm dust shells (D'-type), have very
high rotational velocities, solar metallicities, and are also
s-process over abundant (\citealt{Smi2001}, \citealt{Per2005}).

The vast majority of symbiotic systems, however, seem to contain
normal M-type giants with very little known about their
abundances and no published information about the presence of any
s-process enrichments \citep[][and references therein]{Sch2006}.
Thus far most information about the red giant nature and 
chemistry is derived either from the abundance studies based on
nebular emission lines (\citealt{Nus1988}) or (mostly) TiO molecular
absorption bands (\citealt{Mue1999}).  In addition, analysis of the
first-overtone CO absorption features in $K$-band spectra of a dozen
symbiotic systems indicate sub-solar carbon abundances and
$^{12}$C/$^{13}$C $\sim$ 3 to 30  (\citealt{Sch1992}, \citealt{Sch2003},
\citealt{Sch2006}).  All these have indicated that the surveyed
symbiotic giants are indistinguishable from local M giants in
agreement with abundance studies based on nebular emission lines.

\begin{table}
 \centering
  \caption{Journal of spectroscopic observations}\label{T2}
  \begin{tabular}{@{}lcccc@{}}
  \hline
   Object  & Band & Date       & HJD(mid)     & Orbital phase$\,^a$\\
 \hline
           & $H$  & 16.02.2003 & 2452686.8380 & 0.57 \\
 RW\,Hya   & $K$  & 20.04.2003 & 2452749.6295 & 0.74 \\
           & $K$  & 13.12.2003 & 2452986.8656 & 0.38 \\
           & $K_r$& 03.04.2006 & 2453828.6308 & 0.65 \\
 \hline
           & $H$  & 17.02.2003 & 2452687.7566 & 0.02 \\
 SY\,Mus   & $K$  & 20.04.2003 & 2452749.5817 & 0.12 \\
           & $K$  & 13.12.2003 & 2452986.8250 & 0.50 \\
           & $K_r$& 03.04.2006 & 2453828.5767 & 0.84 \\
\hline
\end{tabular}
\begin{list}{}{}
\item[$^a$] Orbital phases calculated for RW\,Hya from orbital ephemerides
of \citet{KeMi1995} or \citet{Sch1996} and for SY\,Mus from orbital ephemerides
of \citet{Dum1999}
\end{list}
\end{table}

Despite the significant role that knowledge of chemical composition
has to play in improving our understanding of the symbiotic systems,
there are few cases in which abundances are fairly well determined.
Analysis of photospheric chemical abundances has been performed
only for a few bright systems: CH\,Cyg - the brightest symbiotic
system \citep{Sch2006}, V2116\,Oph - the symbiotic system with a
neutron star \citep{Hin2006}, and the symbiotic recurrent novae
T\,CrB and RS\,Oph \citep{Wal2008}.  In all these cases a solar or
nearly solar metallicity was found with Li enhancements 
in RS\,Oph and T\,CrB the only peculiarity. For CH\,Cyg \citet{Sch2006}
found that the isotopic ratios of $^{12}$C/$^{13}$C and $^{16}$O/$^{17}$O are
close to the mean values for single M giants that have experienced
the first dredge-up.  The relatively high metallicity of CH\,Cyg
accounts for the absence of chemical peculiarities similar to those
seen in Ba stars.  The analysis of \citet{Sch2006} has also revealed
significant differences between the C/N and O/N ratios derived from
nebular emission lines versus those from photospheric absorption
lines.   This indicates that an analysis based on emission lines
can, for some circumstances, seriously overestimate the N abundance.
A direct high resolution spectroscopic determination of the
photospheric chemical abundances of red symbiotic giants is needed
to settle this issue and to explore the abundances of more members
of the symbiotic family.

We have undertaken an extensive research program to perform detailed
chemical composition analysis of a sample of over 30 symbiotic
systems.  The analysis is based on near-$IR$ spectra observed at
high-resolution (R $\sim$ 50000) and high S/N ($\sim$100) during
the years 2003-2006 with the Phoenix spectrometer on the Gemini
South telescope.  The spectral regions observed are located in the
$H$ and $K$ photometric bands.  Spectrum synthesis employing standard
LTE analysis and current model atmospheres was used to determine
abundances of CNO and elements around the iron peak (Sc, Ti, Fe,
and Ni) in the stellar photospheres.  We expect our study to provide
clues to resolve the metallicity problem in symbiotic systems as
well as important information for understanding the history of these
systems.

In this paper we present the first analysis of the photospheric
chemical abundances (CNO and elements around iron peak: Sc, Ti, Fe,
and Ni) for two classical S-type symbiotic systems, RW\,Hya and
SY\,Mus.  Both of these systems have well known basic parameters
(Table\,\ref{T1}) with effective temperatures estimated from the
near-IR colors.   Both systems also have well established orbital
solutions with circular orbits.  Near-IR light curves of both systems
show ellipsoidal variations which permit an estimation of their red
giant radii \citep{Rut2007}.  These two systems have tidally distorted
giants similar to those in the classical Z\,And-type symbiotic
systems but unlike these systems they do not show outburst activity.

\begin{table}
\centering
\caption{Velocity parameters$\,^a$ of the cool components
obtained via cross-correlation technique}
\label{TvsiC}
  \begin{tabular}{@{}lccc@{}}
  \hline
                & \multicolumn{3}{c}{RW\,Hya}                 \\
 
 \hline
                & $(V_{\rmn{rot}}^2 \sin^2{i} + \xi^2_{\rmn{t}})^{0.5}$ & $V_{\rmn{rad}}$ & Orbital phase \\
 Feb\,2003      & 5.74                                        & 11.13 $\pm$ 0.15 & 0.57  \\
 Apr\,2003      & 6.63                                        &~~3.04 $\pm$ 0.34 & 0.74  \\
 Dec\,2003      & 6.25                                        & 18.44 $\pm$ 0.32 & 0.38  \\
 Apr\,2006      & 5.57                                        &~~5.33 $\pm$ 0.41 & 0.65  \\
 \hline
                & \multicolumn{3}{c}{SY\,Mus}                 \\
 
 \hline
                & $(V_{\rmn{rot}}^2 \sin^2{i} + \xi^2_{\rmn{t}})^{0.5}$ & $V_{\rmn{rad}}$ & Orbital phase \\
 Feb\,2003      & 3.88                                       & 19.39 $\pm$ 0.15 & 0.02  \\
 Apr\,2003      & 5.98                                       & 19.66 $\pm$ 0.29 & 0.12  \\
 Dec\,2003      & 6.84                                       & 13.14 $\pm$ 0.41 & 0.50  \\
 Apr\,2006      & 5.01                                       &~~4.42 $\pm$ 0.37 & 0.84  \\
  \hline

\end{tabular}
\begin{list}{}{}
\item[$^{a}$]\,Units $\rmn{km}\,\rmn{s}^{-1}$
\end{list}
\end{table}

\begin{table}
\centering
\caption{Quadrature sums of the projected rotational velocities and
microturbulence $(V_{\rmn{rot}}^2 \sin^2{i} + \xi^2_{\rmn{t}})^{0.5}$
from $K$-band \mbox{Ti\,{\sc i}}, \mbox{Fe\,{\sc i}} and \mbox{Sc\,{\sc i}} lines$\,^a$}\label{TvsiF}
\begin{tabular}{@{}lcc@{}} \hline
		& RW\,Hya     & SY\,Mus \\
\hline
 Apr\,2003      & 6.73 $\pm$ 0.60 & 6.91 $\pm$ 0.33 \\ 
 Dec\,2003      & 6.35 $\pm$ 0.24 & 6.97 $\pm$ 0.33 \\ 
 both$\,^{b}$ & 6.54 $\pm$
 0.32 & 6.94 $\pm$ 0.22 \\ \hline
\end{tabular} 
\begin{list}{}{} 
\item[$^{a}$]\,Units $\rmn{km}\,\rmn{s}^{-1}$
\item[$^{b}$]\,Used for synthetic spectra calculations  
\end{list}
 \end{table}

%==================================================================

\section[]{Observations and data reduction}

Spectra of RW\,Hya and SY\,Mus were observed at high-resolution ($R =
\lambda/\Delta\lambda \sim 50000$) and high S/N ratio ($\ga$\,100) in the
near-IR using the Phoenix cryogenic echelle spectrometer on 8\,m Gemini
South telescope.  For both objects one spectrum was observed at
1.56$\,\mu$m ($H$-band) during February 2003, two spectra at
2.23$\,\mu$m ($K$-band) during April and December 2003, and one
spectrum at 2.36$\,\mu$m ($K_{\rm r}$-band) during April 2006.  All the
spectra cover a narrow spectral range of $\sim$100\AA.  The spectra were
extracted and wavelength calibrated using standard reduction techniques
\citep{Joy1992} and all were heliocentric corrected.  In all cases telluric
lines were either not present in the interval observed or were removed by
reference to a hot standard star.  For all spectra the Gaussian instrumental
profile is about of $6\,\rmn{km}\,\rmn{s}^{-1}$ FWHM, corresponding to
instrumental profile widths of 0.31\,\AA\ , 0.44\,\AA\, and 0.47\,\AA\ for
the $H$-band, $K$-band, and $K_{\rm r}$-band spectra, respectively.  The
journal of our spectroscopic observations is given in
Table\,\ref{T2} and the spectra of RW\,Hya and SY\,Mus are shown in
Figures \ref{FRWSc1_H}--\ref{FRWSc1_Kb} and \ref{FSYSc2_H}--\ref{FSYSc2_Kb}, respectively.

\begin{table}
\centering
\caption{Calculated abundances and relative abundances$\,^a$,
velocity parameters$\,^b$, and uncertainties$\,^c$ for RW\,Hya and
SY\,Mus
}\label{TaSc1}
  \begin{tabular}{@{}lccccr@{}}
  \hline
         &\multicolumn{2}{c}{RW\,Hya} &\multicolumn{2}{c}{SY\,Mus} &\\
  $X$    & $\log{\epsilon(X)}$ & [$X$]         & $\log{\epsilon(X)}$  & [$X$]  & n$\,^d$\\
  \hline
  $^{12}$C         & 7.53$\pm$0.02  & -0.90$\pm$0.07 & 8.17$\pm$0.01& -0.26$\pm$0.06  & 90\\
  N                & 7.46$\pm$0.03  & -0.37$\pm$0.08 & 8.11$\pm$0.02& +0.28$\pm$0.07  & 62\\
  O                & 8.17$\pm$0.01  & -0.52$\pm$0.06 & 8.66$\pm$0.01& -0.03$\pm$0.06  & 49\\
  Sc               & 2.71$\pm$0.05  & -0.44$\pm$0.09 & 3.97$\pm$0.05& +0.82$\pm$0.09  &  1\\
  Ti               & 4.49$\pm$0.05  & -0.46$\pm$0.10 & 5.12$\pm$0.03& +0.17$\pm$0.08  &  9\\
  Fe               & 6.74$\pm$0.02  & -0.76$\pm$0.06 & 7.42$\pm$0.02& -0.08$\pm$0.06  & 21\\
  Ni               & 5.63$\pm$0.03  & -0.59$\pm$0.07 & 6.37$\pm$0.03& +0.15$\pm$0.07  &  3\\
  $^{12}$C/$^{13}$C& 6$\pm$2~ & ... & 10$\pm$3~ & ...             & $\frac{66}{16}^e$\\
  \hline
$\xi_{\rmn{t}}$         &\multicolumn{2}{c}{1.8$\pm$0.2} &\multicolumn{2}{c}{2.0$\pm$0.2} & ... \\
$V_{\rmn{rot}} \sin{i}$ &\multicolumn{2}{c}{6.3$\pm$0.2} &\multicolumn{2}{c}{6.6$\pm$0.2} & ... \\
  \hline
\end{tabular}
\begin{list}{}{}
\item[$^a$] Relative to the Sun [$X$] abundances estimated in relation to the solar composition of \citet{Asp2009}
\item[$^b$] Units $\rmn{km}\,\rmn{s}^{-1}$ 
\item[$^c$] 3$\sigma$
\item[$^d$] Number of lines used 
\item[$^e$] Number of lines used to estimate $^{12}$C/$^{13}$C isotopic ratio:  66 lines for  $^{12}$C and 16 lines for $^{13}$C.\\
\end{list}
\end{table}

%==================================================================

\section{Analysis and results}\label{secanalysis}

The observed $H$-band region is dominated by strong molecular lines
from the CO second overtone, OH first overtone, and relatively weak
CN red systems lines.  It also contains numerous atomic absorption
lines mainly from \mbox{Fe\,{\sc i}} with a few \mbox{Ti\,{\sc i}} and \mbox{Ni\,{\sc i}} lines.  The
$K$-band region is dominated by atomic \mbox{Ti\,{\sc i}}, \mbox{Fe\,{\sc i}}, and \mbox{Sc\,{\sc i}}
lines.  There are also numerous weak CN molecular lines.  \mbox{Ca\,{\sc i}}
and \mbox{Ba\,{\sc i}} atomic lines are also present but too weak to be useful
in chemical abundance analysis.  $K_{\rm r}$-band region has strong
first overtone CO molecular absorption lines coming from transitions
quite far from the band head.  This results in slight problems in
reliably setting the continuum.  While dominated by lines from
$^{12}$CO the presence of lines from the $^{13}$CO isotopologue
(mainly from the 2--0 transition) make this spectrum useful for 
evaluating the $^{12}$C/$^{13}$C isotopic ratio.  There are also
three atomic lines, two from \mbox{Fe\,{\sc i}} and one from \mbox{Sc\,{\sc i}}.

The analysis was performed by fitting synthetic spectra to the
observed spectra using a similar method to that used by \citet{Sch2006}
in evaluating the CH\,Cyg abundances.  The calculations of synthetic
spectra over the entire observed spectral region were performed
using the code WIDMO developed by M.R. Schmidt \citep{Sch2006}.
To perform the $\chi^{2}$ minimization a special overlay was
performed on the code WIDMO with use of the simplex algorithm.  This
procedure used the method described in \citet{Bra1998} and the
flowchart of \citet{KaMi1999}.  To evaluate quantitatively the fit
quality, the residuals were calculated during each iteration, by
differencing the synthesized and observed spectra at each point in
the used spectral range.  The simplex algorithm searched for the
minimum of the standard deviation of the residuals to find an optimal
fit.

For elemental abundance determination stellar effective temperature ($T_{\rm
eff}$) and surface gravity ($\log{g}$) are required as input parameters.  In
addition information about micro- and macro-turbulences, line
identifications with values of quantities that characterize the transitions
and the models of the atmospheres are needed.  MARCS model atmospheres developped by
\citet{Gus2008} were used in our calculations.  The input effective
temperatures ($T_{\rm eff}$) were estimated from the known spectral types,
M2 and M5 \citep[][and references therein]{Bel2000}, adopting the
calibration by \citet{Ric1999}.  The resulting values are $T_{\rm eff} =
3655\pm80$\,K for RW\,Hya and $T_{\rm eff} = 3355\pm75$\,K for SY\,Mus.  A
$(J-K)$ color of 1.15 and 1.40 \citep{Rut2007} and color excesses $E(B-V)$ =
0.02--0.10 and 0.4--0.5 results in intrinsic colors of $(J-K)_0$ $\sim$1.1
and $\sim$1.2 for RW\,Hya and SY\,Mus.  Using the \citet{Kuc2005} $T_{\rm
eff}$--$\log{g}$--color relation for late-type giants we can estimate the
effective temperatures as 3600--3700\,K and $\lesssim$3500\,K, respectively,
in good agreement with the above values.  The surface gravities from this
relation are $\log{g}\approx 0.5$ in both cases, 
in agreement with the values derived from the published red giant parameters  
(Table\,\ref{T1}), $\log{g}\approx 0.5$ and $\log{g}\approx 0.3$ for RW Hya and SY Mus, respectively.  
As a result the model atmospheres used had $\log{g} = 0.5$, solar chemical composition
([Fe/H] = 0.0), and effective temperatures of $T_{\rm eff} = 3700$\,K
and $T_{\rm eff} = 3400$\,K for RW\,Hya and SY\,Mus, respectively.
In the case of RW Hya our analysis revealed significantly sub-solar  metallicity, [Fe/H]$\sim$-0.7. 
The whole analysis was therefore repeated adopting the model atmosphere with [Fe/H]=-0.75.
\begin{table}
\centering
\caption{Calculated abundances and relative abundances$\,^a$ for
RW\,Hya and SY\,Mus  where all radial and rotational parameters were set as a free
parameters}\label{TaM}

  \begin{tabular}{@{}lccccr@{}}
  \hline
         &\multicolumn{2}{c}{RW\,Hya} &\multicolumn{2}{c}{SY\,Mus} &\\
  $X$    & $\log{\epsilon(X)}$ & [$X$]         & $\log{\epsilon(X)}$  &
[$X$]  & n$\,^b$\\
  \hline
  $^{12}$C         & 7.54 & -0.89 & 8.10 & -0.33 & 90\\
  N                & 7.46 & -0.37 & 8.11 & +0.28 & 62\\
  O                & 8.17 & -0.52 & 8.58 & -0.11 & 49\\
  Sc               & 2.75 & -0.40 & 4.02 & +0.87 &  1\\
  Ti               & 4.54 & -0.41 & 5.13 & +0.18 &  9\\
  Fe               & 6.75 & -0.75 & 7.39 & -0.11 & 21\\
  Ni               & 5.63 & -0.59 & 6.36 & +0.14 &  3\\
  $^{12}$C/$^{13}$C& 5.0  & ...   & 8.0  & ...   & $\frac{66}{16}^{\star}$\\
  \hline
\end{tabular}
\begin{list}{}{}
\item[$^a$] Relative to the Sun [$X$] abundances estimated in relation to
the solar composition of \citet{Asp2009} 
\item[$^b$] Number of lines used, $^{12}$C/$^{13}$C number as in Table 5
\end{list}
\end{table}

\begin{table}
 \centering
\caption{Velocity parameters$\,^a$ of the cool giants where radial and rotational parameters were set as a free
parameters}\label{TvsiM}

  \begin{tabular}{@{}lccc@{}}
  \hline
                & \multicolumn{3}{c}{RW\,Hya}                 \\%&
 
 \hline
$\xi_{\rmn{t}}$ &  \multicolumn{3}{c}{1.74 $\pm$ 0.27}        \\%&
 \hline
                & $V_{\rmn{rot}} \sin{i}$ & $V_{\rmn{rad}}$  & Orbital phase \\
 Feb\,2003      & 6.0~~$\pm$ 0.4~~        & 11.09 $\pm$ 0.47 & 0.57  \\
 Apr\,2003      & 6.6~~$\pm$ 0.7~~        &~~2.85 $\pm$ 0.59 & 0.74  \\
 Dec\,2003      & 7.1~~$\pm$ 0.5~~        & 18.77 $\pm$ 0.31 & 0.38  \\
 Apr\,2006      & 8.9~~$\pm$ 0.5~~        & ~5.06 $\pm$ 0.35 & 0.65  \\
 \hline
                & \multicolumn{3}{c}{SY\,Mus}                 \\
 
 \hline
$\xi_{\rmn{t}}$ &  \multicolumn{3}{c}{1.95 $\pm$ 0.30}       \\
 \hline
                & $V_{\rmn{rot}} \sin{i}$ & $V_{\rmn{rad}}$  & Orbital phase \\
 Feb\,2003      & 5.4~~$\pm$ 0.5~~        & 19.28 $\pm$ 0.26 & 0.02  \\
 Apr\,2003      & 5.9~~$\pm$ 0.6~~        & 19.77 $\pm$ 0.38 & 0.12  \\
 Dec\,2003      & 6.6~~$\pm$ 0.5~~        & 13.26 $\pm$ 0.36 & 0.50  \\|
 Apr\,2006      & 6.5~~$\pm$ 0.6~~        &~~4.40 $\pm$ 0.20 & 0.84  \\
  \hline

\end{tabular}
\begin{list}{}{}
\item[$^a$] Units $\rmn{km}\,\rmn{s}^{-1}$ 
\end{list}
\end{table}

We used a number of standard sources to characterize the spectral lines. 
For the macroturbulence velocity, $\zeta_{\rmn{t}}$, we used
$3\,\rmn{km}\,\rmn{s}^{-1}$, typical for cool red giants \citep[see for
example][]{Fek2003}.  The atomic data were taken from the VALD database
\citep{Kup1999} in the case of $K$- and $K_{r}$-bands and from the list
given by \citet{MeBa1999} in the $H$-band.  The vacuum wavelengths,
excitation potentials, and $gf$-values for the vibration-rotation lines of
CO isotopes are adopted from \citet{Goo1994}. The lists of $^{12}$CN,
and OH molecular lines were adopted from the database of
\citet{Kur1999}. The complete list of the lines identified and selected for
our abundance analysis, with the excitation potentials ($EP$) and
$gf$-values for transitions, is shown in Tables\ref{TLA} and \ref{TLM} in
the Appendix.

To fit synthetic spectra to observed spectra it is necessary to
find, in addition to the parameters of chemical composition (C, N,
O, Sc, Ti, Fe, Ni in this case), values for a number of velocities.
These include the microturbulent velocity, $\xi_{\rmn t}$, the
radial velocity, $V_{\rmn{rad}}$, and the rotational velocity,
$V_{\rmn{rot}}$ $\sin{i}$.  The adopted strategy relies on handling
the three spectral regions in the $H$- and $K$-band ranges
as if they were a single spectrum to obtain a solution for the
chemical composition.  The chemical composition should not depend
on the orbital phase.  One set of the abundance parameters and
microturbulent velocity will apply to all spectra of a given object.
On the other hand, the radial velocities $V_{\rmn{rad}}$ depend on
the orbital phase and the rotational velocities $V_{\rmn{rot}}$
$\sin{i}$ can also depend on tidal forces in the outer regions of
the giant.

To obtain radial and rotational velocities we used a cross-correlation
technique similar to that of \citet{Car2011}. The method enables
the evaluation of rotational velocities in spectra rich with lines
where unblended lines are rare, as is the case of our $H$- and
$K_{\rmn{r}}$-band spectra. For templates we used synthetic spectra
generated from the line list of Appendix\,A (Tables\ref{TLA} and
\ref{TLM}) broadened only by the instrumental profile. We measured
the FWHM of the cross-correlation peak, $\mu_{\star}$, for each
observed spectrum and the FWHM of auto-correlation peak, $\mu_{\rmn{t}}$,
for each appropriate template using IRAF's $fxcor$ task. The widths
of spectral lines attributable to stellar processes were obtained
by subtraction of the contributions from the template and instrumental
profile, $\omega_{\star} = (\mu_{\star}^2 - 0.5 \mu_{\rmn{t}}^2 -
\omega_{\rmn{i}}^2)^{0.5}$. Next the quantity $\omega_{\star}$ was
converted to a so called total stellar broadening ($\beta_{\rmn{Gray}}$)
using a second order polynomial applicable to FWHM measurements made
at any wavelength in velocity units \citep[eq.\,1 in ][]{Car2011}:
$\omega_{\star} (\rmn{km s}^{-1}) = 1.89582 + 1.16526 \beta_{\rmn{Gray}}
+ 0.0065 \beta_{\rmn{Gray}}^2$.  Finally we obtained the quadrature
sums of the projected rotational velocities and microturbulence:
$(V_{\rmn{rot}}^2 \sin^2{i} + \xi^2_{\rmn{t}})^{0.5} =
(\beta_{\rmn{Gray}}^2 - \zeta^2_{\rmn{t}})^{0.5}$.  The velocity
parameters derived are shown in Table\ref{TvsiC}.

Our $K$-band spectra contain a few relatively strong atomic lines
that are not blended.  These lines enable a direct measurement of
the FWHMs of the stellar line profiles. \citet{Fek2003} used three
\mbox{Ti\,{\sc i}}, two \mbox{Fe\,{\sc i}}, and one \mbox{Sc\,{\sc i}} line to measure rotational
velocities in a dozen symbiotic systems. We used the same lines
to measure rotational velocities in RW\,Hya and SY\,Mus based on
the total stellar broadening $\beta_{\rmn{Gray}}$ for each spectrum
separately and for 12 lines from both spectra simultaneously. To
measure FWHMs the IRAFs $splot$ tool have been used. The results
are shown in Table\,\ref{TvsiF}. The values are in agreement within
the uncertainties with those obtained via another method (see
Tables\,\ref{TvsiC} and \ref{TvsiM}).

To perform solutions using the simplex algorithm $n+1$, $n$ dimensional
sets of free parameters must be prepared that form a grid in parameter
space called simplex (see e.g.  \citealt{KaMi1999}; \citealt{Bra1998}).
We have 7 free parameters for abundances. Two additional free
parameters are the projected rotational velocity ($V_{\rmn{rot}}\sin{i}$)
and microturbulence ($\xi_{\rmn{t}}$).  The quadrature sum of these
quantities is treated as a constant adopted from the previously
obtained value (see Table\,\ref{TvsiF}). To estimate the input
abundance parameters the adopted procedure was as follows.  As a
first approximation solar composition \citep{Asp2009} was adopted
and the oxygen abundance was found by fitting (by eye) the OH lines
in the $H$-band spectrum.  The carbon abundance was adjusted next
to find the best fit to the CO lines.  Following this the nitrogen
abundance was adjusted by fitting the CN lines in $H$- and $K$-band
spectra.  The last step was to fit the atomic lines. The procedure
was repeated for several iterations. In this way the initial values
of the free parameters were found around which the simplexes were
build.

\begin{figure}
\resizebox{\hsize}{!}{\includegraphics{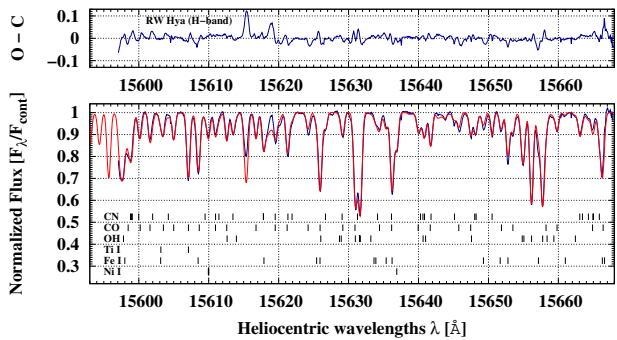}}
\caption{The spectrum of RW\,Hya observed in February 2003 (thin
line) in the $H$-band and a synthetic spectrum (thick line) calculated
using the final abundances and $^{12}$C/$^{13}$C isotopic ratio
(Table\,\ref{TaSc1}). Molecular (OH, CO, CN) and atomic (\mbox{Ti\,{\sc i}},
\mbox{Fe\,{\sc i}}, \mbox{Ni\,{\sc i}}) lines used in the solution of the chemical composition
are identified by ticks.}\label{FRWSc1_H}
\end{figure}

\begin{figure}
\resizebox{\hsize}{!}{\includegraphics{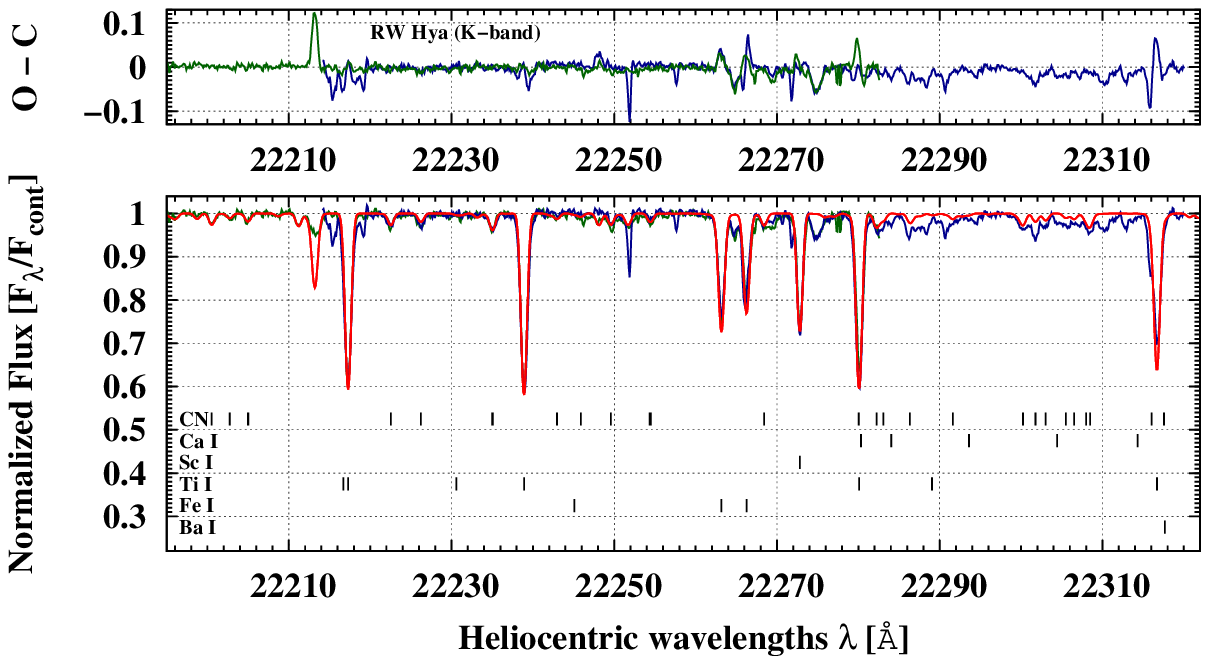}}
\caption{Spectra of RW\,Hya observed in April 2003 (thin solid line)
and December 2003 (thin dot-dashed line) in the $K$-band and a synthetic
spectrum (thick solid line) calculated using the final abundances
and $^{12}$C/$^{13}$C isotopic ratio (Table\,\ref{TaSc1}). Molecular
(CN) and atomic (\mbox{Sc\,{\sc i}}, \mbox{Ti\,{\sc i}}, \mbox{Fe\,{\sc i}}) lines used in the solution
of the chemical composition are identified by ticks.}\label{FRWSc1_K}
\end{figure}

\begin{figure}
\resizebox{\hsize}{!}{\includegraphics{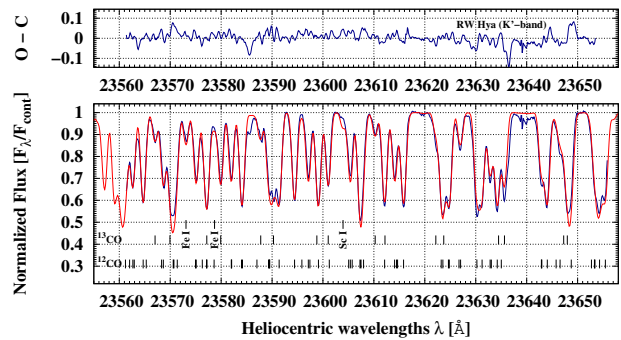}} 
\caption{The spectrum of RW\,Hya observed in April 2006 (thin line)
in the $K_{\rm r}$-band and a synthetic spectrum (thick line) calculated
using the final abundances and $^{12}$C/$^{13}$C isotopic ratio
(Table\,\ref{TaSc1}). Molecular ($^{12}$CO, $^{13}$CO) and atomic
(\mbox{Sc\,{\sc i}}, \mbox{Fe\,{\sc i}}) lines are identified by ticks.}\label{FRWSc1_Kb}
\end{figure}

\begin{figure} 
\resizebox{\hsize}{!}{\includegraphics{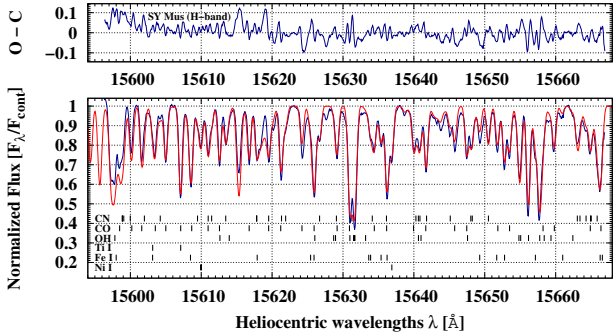}} 
\caption{The spectrum of SY\,Mus observed in February 2003 in the
$H$-band (thin line) and a synthetic spectrum (thick line) calculated
using the final abundances and $^{12}$C/$^{13}$C isotopic ratio
(Table\,\ref{TaSc1}). Molecular (OH, CO, CN) and atomic (\mbox{Ti\,{\sc i}},
\mbox{Fe\,{\sc i}}, \mbox{Ni\,{\sc i}}) lines used in the solution of the chemical composition
are identified by ticks.}\label{FSYSc2_H}
\end{figure}

\begin{figure}
\resizebox{\hsize}{!}{\includegraphics{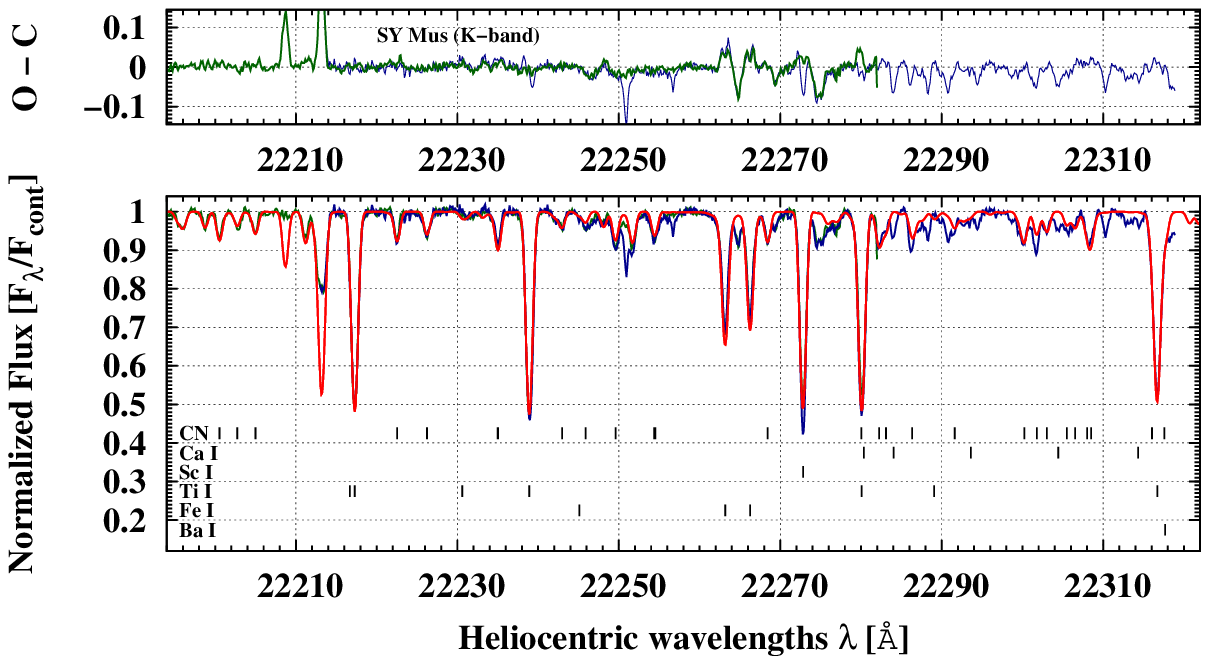}} 
\caption{Spectra of SY\,Mus observed in April 2003 (thin solid line)
and December 2003 (thin dot-dashed line) in the $K$-band and a
synthetic spectrum (thick solid line) calculated using the final
abundances and $^{12}$C/$^{13}$C isotopic ratio (Table\,\ref{TaSc1}).
Molecular (CN) and atomic (\mbox{Sc\,{\sc i}}, \mbox{Ti\,{\sc i}}, \mbox{Fe\,{\sc i}}) lines used in the
solution of the chemical composition are identified by
ticks.}\label{FSYSc2_K}
\end{figure}

\begin{figure}
\resizebox{\hsize}{!}{\includegraphics{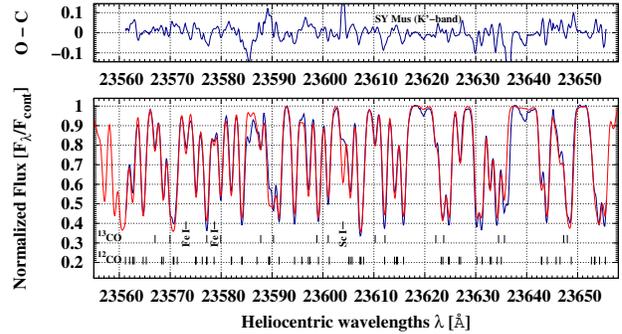}}
\caption{The spectrum of SY\,Mus observed in April 2006 (thin line)
in the $K_{\rm r}$-band and a synthetic spectrum (thick line) calculated
using the final abundances and $^{12}$C/$^{13}$C isotopic ratio and
$^{12}$C/$^{13}$C isotopic ratio (Table\,\ref{TaSc1}). Molecular
($^{12}$CO, $^{13}$CO) and atomic (\mbox{Sc\,{\sc i}}, \mbox{Fe\,{\sc i}}) lines are identified
by ticks.}\label{FSYSc2_Kb} 
\end{figure}

\begin{figure}
\resizebox{\hsize}{!}{\includegraphics{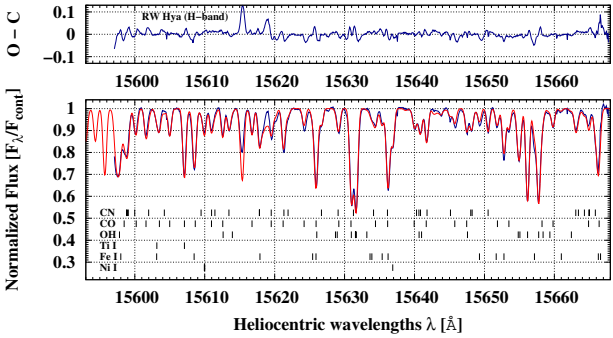}}
\caption{The spectrum of RW\,Hya observed in February 2003 in the
$H$-band (thin line) and a synthetic spectrum (thick line) calculated
using the final abundances and $^{12}$C/$^{13}$C isotopic ratio
(Table\,\ref{TaM}). This is the case where all radial and rotational
parameters were allowed to be free parameters for the solution. Molecular
(OH, CO, CN) and atomic (\mbox{Ti\,{\sc i}}, \mbox{Fe\,{\sc i}}, \mbox{Ni\,{\sc i}}) lines used in the
solution of the chemical composition are identified by ticks.}\label{FRW_H}
\end{figure}

\begin{figure}
\resizebox{\hsize}{!}{\includegraphics{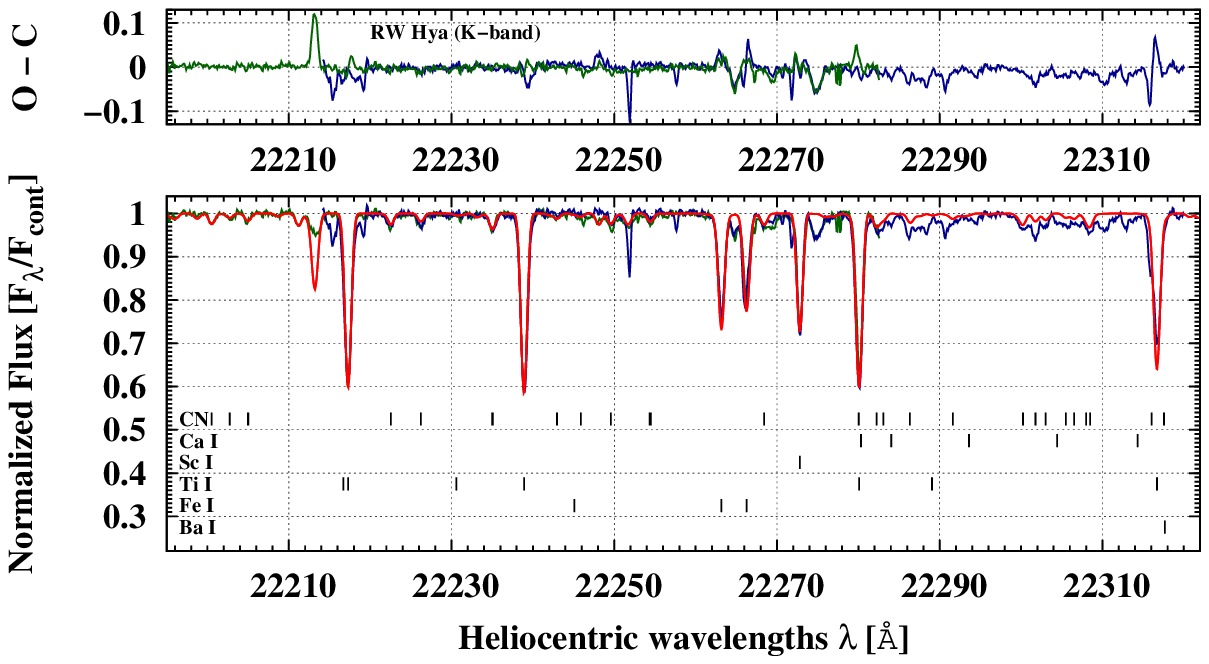}}
\caption{Spectra of RW\,Hya observed in April 2003 (thin solid line)
and December 2003 (thin dot-dashed line) in the $K$-band and a
synthetic spectrum (solid line) calculated using the final abundances
and $^{12}$C/$^{13}$C isotopic ratio (Table\,\ref{TaM}). This is
the case where all radial and rotational were allowed to be free
parameters for the solution. Molecular (CN) and atomic (\mbox{Sc\,{\sc i}},
\mbox{Ti\,{\sc i}}, \mbox{Fe\,{\sc i}}) lines used in the solution of the chemical composition
are identified by ticks.}\label{FRW_K} 
\end{figure}

\begin{figure}
\resizebox{\hsize}{!}{\includegraphics{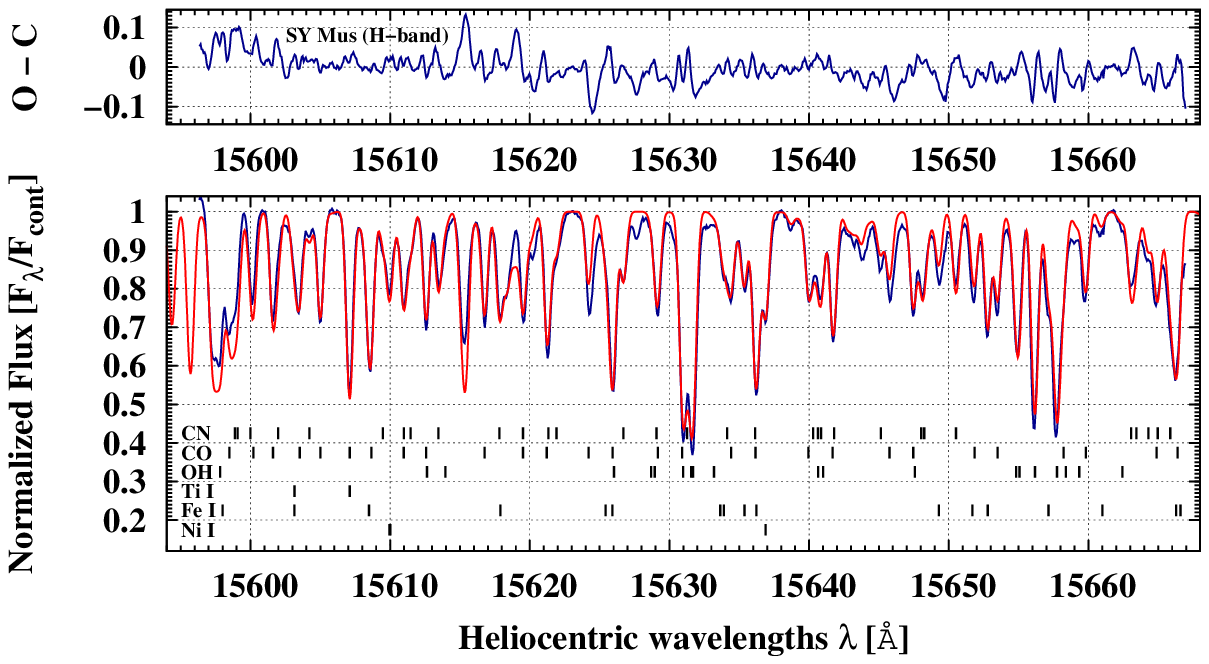}}
\caption{The spectrum of SY\,Mus observed in February 2003 in the
$H$-band (thin line) and a synthetic spectrum (thick line) calculated
using the final abundances and $^{12}$C/$^{13}$C isotopic ratio
(Table\,\ref{TaM}). This is the case where all radial and rotational
parameters were allowed to be free parameters for the solution.
Molecular (OH, CO, CN) and atomic (\mbox{Ti\,{\sc i}}, \mbox{Fe\,{\sc i}}, \mbox{Ni\,{\sc i}}) lines used
in the solution of the chemical composition are identified by
ticks.}\label{FSY_H} 
\end{figure}

\begin{figure}
\resizebox{\hsize}{!}{\includegraphics{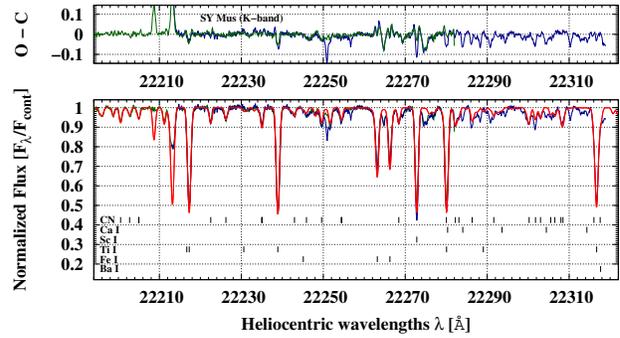}}
\caption{Spectra of SY\,Mus observed in April 2003 (thin solid line)
and December 2003 (thin dot-dashed line) in the $K$-band and a
synthetic spectrum (solid line) calculated using the final abundances
and $^{12}$C/$^{13}$C isotopic ratio (Table\,\ref{TaM}). This is
the case where all radial and rotational parameters were allowed
to be free parameters for the solution. Molecular (CN) and atomic
(\mbox{Sc\,{\sc i}}, \mbox{Ti\,{\sc i}}, \mbox{Fe\,{\sc i}}) lines used in the solution of the chemical
composition are identified by ticks.}\label{FSY_K} 
\end{figure}

Nine different initial simplexes were used in both the RW\,Hya and
SY\,Mus cases with each repeated with different values of the
microturbulent velocity $\xi_{\rmn t}$. Although the microturbulent
velocity is a free parameter in our solution it is not an element
of the simplex.  Rather the best value was searched for in the
1.2--2.6\,$\rmn{km}\,\rmn{s}^{-1}$ range with use of golden ratio
method.  After we found the sets of parameters that give the best
fit to $H$- and $K$-band spectra we applied the derived chemical
abundances to the $K_{\rmn r}$ spectrum as a fixed parameter and
we searched for the $^{12}$C/$^{13}$C ratio with the rotational
velocity as an additional free parameter.  After obtaining the
optimal fit we made a reconciliation of $^{12}$C abundance and
$^{12}$C/$^{13}$C isotopic ratio within 3 iterations.  The final
abundances for CNO elements and atomic lines (\mbox{Sc\,{\sc i}}, \mbox{Ti\,{\sc i}}, \mbox{Fe\,{\sc i}},
\mbox{Ni\,{\sc i}}) in the scale of $\log{\epsilon}(X) = \log{N(X) N(H)^{-1}} +
12.0$, the isotopic ratios $^{12}$C/$^{13}$C, microturbulences and
projected rotational velocities are summarized in Table\,\ref{TaSc1}
together with the formal uncertainties. Our final synthetic spectra
fitted to the observed ones are show in
Figures\,\ref{FRWSc1_H}--\ref{FSYSc2_Kb}.  Additionally, we obtained
independent solutions for each spectrum for the case when all radial
and rotational velocities are treated as a free parameters.  The
resulting abundances are shown in Table\ref{TaM} and the velocity
parameters are shown in Table\,\ref{TvsiM}, and the final synthetic spectra fitted to the observed ones are presented in Figures\,\ref{FRW_H}--\ref{FSY_K}.

\begin{table*}
 \centering
  \caption{Sensitivity of abundances to uncertainties in the stellar parameters}\label{TdSc1}
  \begin{tabular}{@{}lccccccc@{}}
  \hline
  $\Delta X$ & \multicolumn{2}{c}{$\Delta T_{\rmn{eff}} = +100$\,K} & \multicolumn{2}{c}{$\Delta \log{g} = +0.5$} & \multicolumn{2}{c}{$\Delta \xi_{\rmn{t}} = +0.5$}  & $\Delta [Fe/H] = +0.25$ \\
                 & RW\,Hya & SY\,Mus  & RW\,Hya & SY\,Mus & RW\,Hya & SY\,Mus  & RW Hya\\
  \hline
  C              & +0.04    & +0.03   & +0.21  & +0.22    & -0.01   & -0.07  & 0.03\\
  N              & +0.08  & +0.04   & -0.05  & +0.02  & -0.08  & -0.12  & 0.16 \\
  O              & +0.16   & +0.12  & +0.03 & +0.08  & -0.06  & -0.10 & 0.14\\
  Sc             & +0.15  & +0.12  & +0.08 & +0.15   & -0.15 & -0.61 & 0.03\\
  Ti             & +0.12  & +0.08 & +0.12  & +0.15  & -0.40  & -0.50 & 0.07 \\
  Fe             & -0.01   & -0.04   & +0.11 & +0.15  & -0.14  & -0.15  & 0.02 \\
  Ni             & -0.00  & -0.04  & +0.11  & +0.21  & -0.08  & -0.23  & 0.01 \\
  \hline
\end{tabular}
\end{table*}

We also made additional fits with MARCS atmosphere models varying the
effective temperatures by $\pm$100\,K, and $\log{g}$ and the
microturbulence, $\xi_{\rm t}$, by $\pm$0.5, and in, the case of RW Hya, also the metallicity [Fe/H] by +0.25 to estimate the sensitivity of
abundances to uncertainties in stellar parameters: $T_{\rm eff}$, $\log{g}$,
$\xi_{\rm t}$, for each species.  The changes in the abundance for each
element as a function of each stellar parameter are listed in
Tables\,\ref{TdSc1}.  Uncertainties in the stellar parameters
have the strongest impact on the accuracy of the obtained chemical
composition.

The abundance of scandium is based only on one strong \mbox{Sc\,{\sc i}} line
at $\lambda \sim 22272.8$\AA\ and it may be less reliable than other
abundances. There are two another scandium \mbox{Sc\,{\sc i}} lines present in
$K$-band spectrum at $\lambda \sim 22213.2$\AA\ and $\lambda \sim
22208.7$\AA \,(see Appendix Table\ref{TLA}). However they show quite
different behaviour and when used they indicate an abundance lower
by $\sim$1\,dex. We suspect, the atomic data published for these
transitions are not reliable. The use of other published data
for these transitions (eg. \citet{Peh2012} or \citet{WiFu1975})
leads to almost identical results.  Broadening of the infrared \mbox{Sc\,{\sc i}} lines 
by hyperfine structure \citep{aboussaid_et_al_2006} has not
been included in the analysis.

%==================================================================

\section{Discussion}

To fit the synthetic spectra it is necessary to take into account
the stellar motions (both the radial and rotational velocities) and
also velocity fields below the stellar photosphere
(turbulences). In section\,\ref{secanalysis} we obtained the
rotational and radial velocities via various methods (compare
Tables \ref{TvsiC}, \ref{TvsiF}, and \ref{TvsiM}) as well as microturbulences
(Tables \ref{TaSc1} and \ref{TvsiM}). The microturbulent velocities
$\xi_{\rm t}$ obtained in our solutions are self-consistent and are
consistent with those obtained for Galactic M-type Giants by
(\citealt[][1990]{SmLa1986}) and for LMC red giants by \citet{Smi2002}.

The radial velocities we measured by cross-correlation and fitting
synthetic spectra are perfectly compatible. The values obtained for
RW\,Hya are in accord with synthetic radial velocities predicted
from the orbit of \citet{KeMi1995} with an accuracy generally better
than 1\,$\rmn{km}\,\rmn{s}^{-1}$.  One spectrum observed on Feb\,2003
is in poorer agreement, $\sim$2\,$\rmn{km}\,\rmn{s}^{-1}$.  The
radial velocities for SY\,Mus agree with synthetic
velocities from the orbit of \citet{Dum1999} with accuracy of $\sim$
1--2\,$\rmn{km}\,\rmn{s}^{-1}$.  The exception here is for a spectrum
observed on Feb\,2003 where the difference is large,
$\sim$5\,$\rmn{km}\,\rmn{s}^{-1}$.

In these stars the rotational velocities, $V_{\rm{rot}} \sin{i}$,
give the largest contribution to the physical line broadening. In
the case of spectra with relatively strong atomic lines and weak
molecular lines, like those in $K$-band region, the rotational
velocity obtained by cross-correlation, by fitting synthetic spectra,
and from FWHM of strong unblended lines give perfectly consistent
results. However, velocities obtained for
spectra with strong molecular lines (strongly blended) are considerably
different depending on the method used and appear to be significantly
underestimated by the cross-correlation method.

Our results obtained from the spectra taken at different, random
orbital phases, differ slightly. The origin of these differences
is not clear. On the one hand, if these stars were significantly
tidally distorted, the projected rotational velocities might vary
with orbital phase. On the other hand this could be due to strong
blending of the lines in some spectra (discussed above) or maybe
to the use of an uncertain value for the instrumental profile.
However, in general these differences are only slightly higher than
the limits set by the uncertainties.  The only exception is somewhat
higher rotational velocity derived for RW Hya from spectrum obtained
in Apr 2006 than those resulting from the 2003 data.  
\citet{Fek2003} estimated $V_{\rm{rot}} \sin{i}$ = 5 $\pm$
1\,$\rmn{km}\,\rmn{s}^{-1}$ for RW\,Hya, measured in the $K$-band
region, which is somewhat lower than our values.

Our analysis of the chemical abundances has revealed an approximately solar
metallicity ([Fe/H]$\sim$0.0) for SY\,Mus and significantly sub-solar
metallicity ([Fe/H]$\sim$-0.75) for RW\,Hya.  CNO abundances in SY\,Mus
(Tables\,\ref{TaSc1} and \ref{TaM}) are close to the averaged values for
'normal' single M giants obtained by \citet{SmLa1990}.  In RW\,Hya the
CNO abundances are significantly reduced.

\begin{figure}
\resizebox{\hsize}{!}{\includegraphics{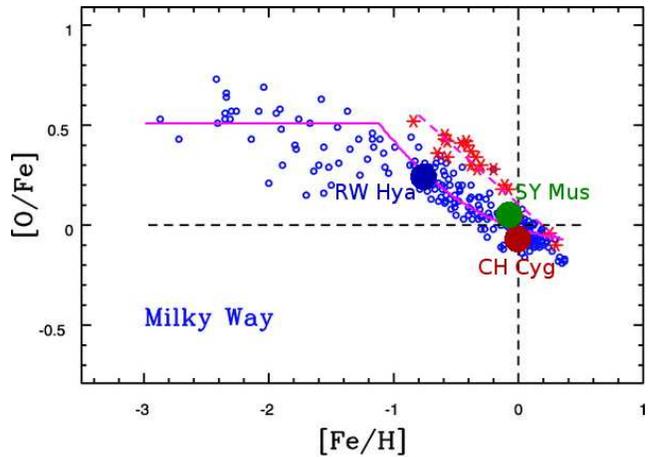}}
\caption{Oxygen relative to iron for various stellar populations
\citep{CuSm2006}. The position of RW\,Hya indicates membership in
the thick disc/halo whereas SY\,Mus and CH\,Cyg are members of the
thin disc.}
\label{O_to_Fe}
\end{figure}

\citet{CuSm2006} showed that different stellar populations are separated in
the [O/Fe]-[Fe/H] plane.  Figure\,\ref{O_to_Fe} shows the position of
RW\,Hya, and SY\,Mus in Fig.  5 of \citet{CuSm2006}.  The symbiotic star CH
Cyg is also plotted with its abundances derived by \citet{Sch2006}.  The
values of [O/Fe] versus [Fe/H] for RW\,Hya locate it among the thick
disk/halo giants whereas those of SY\,Mus, and
CH Cyg are typical for thin disk objects.  These results are consistent with
the location of these objects in the Milky Way, as well as their systemic
velocities and proper motions.  In particular, RW\,Hya has a relatively high
galactic latitude, $b=36.5^\circ$, and appreciable proper motion,
$\mu_{\alpha} \cos \delta=-18.6\pm1.6$\, mas and $\mu_{\delta}=14.1\pm1.6$\,
mas \citep{hog2000}.  Its significant proper motion has been known for a
long time which led \citet{Tif1958} to suggestion that RW\,Hya is a halo
object.  The sub-solar metallicity derived for RW\,Hya is quite remarkable
as it may host a relatively massive, $\sim$0.8 $M_\odot$, white dwarf
\citep{Otu2013} and the present giant could be polluted by s-process
elements from the former TP-AGB companion.  Unfortunately, our infrared
spectra do not cover any s-process element lines.

According to the predictions of the stellar evolution models the
abundance of the $^{13}$C isotope in the photospheres of low mass stars
should increase due to extensive mixing processes
during the ascent of the red giant branch (RGB).  The very low $^{12}$C/$^{13}$C
isotopic ratios obtained for RW\,Hya and SY\,Mus (Tables\,\ref{TaSc1}
and \ref{TaM}) indicate that their giants have experienced first
dredge-up when they were enriched in the $^{13}$C isotope and that
they belong to a population with medium or low metallicity
\citep[see][]{Kel2001}.  \citet{CuSm2006} adapted C and N abundances
for monitoring the dredge-up on the RGB in the Galactic bulge. In
Figure 12
$^{12}$C and $^{14}$N abundances are plotted for RW\,Hya and SY\,
Mus as well as those for CH\,Cyg from \citet[][Fig. 4]{Sch2006}.
The location of RW\,Hya, and SY\, Mus overlap with the bulge red
giants in the $^{14}$N-rich region whereas CH\, Cyg remains near
the solar line.  The conclusion from of this plot is that the CN
mixing has been more extensive in RW\,Hya and SY\,Mus than in
CH\,Cyg. This is also reflected in the lower  isotopic ratios,
$^{12}$C/$^{13}$C=6 and 8,  in the former two objects as compared
with $^{12}$C/$^{13}$C=18 in CH\,Cyg \citep{Sch2006}.

\begin{figure}
\resizebox{\hsize}{!}{\includegraphics{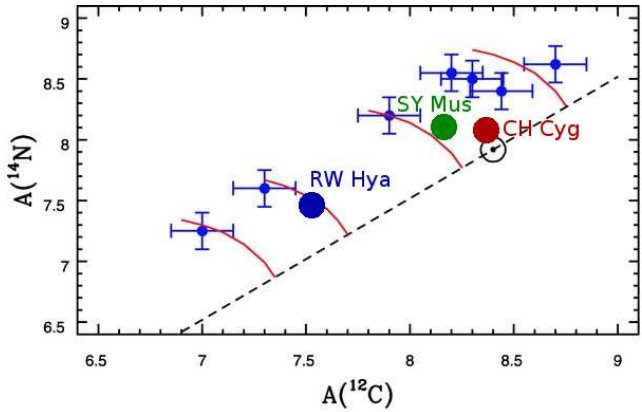}}
\caption{Nitrogen relative carbon for the stars from \citet{CuSm2006}
and our targets. The dashed line represents scaled solar
abundances, [$^{12}$C/Fe]=0 and [$^{14}$N/Fe]=0, whereas the solid
curves delineate constant $^{12}$C+$^{13}$C `CN mixing
lines'. }

\label{N_to_C}
\end{figure}

%==================================================================

\section{Conclusions}

We have performed a detailed analysis of the photospheric abundances of CNO
and elements around the iron peak (Sc, Ti, Fe, and Ni) for the red giant
components of RW Hya and SY Mus.  Our analysis reveals a significantly
sub-solar metallicity ([Fe/H]$\sim$\mbox{-0.75}) for the RW\,Hya giant,
confirming its membership in the Galactic halo population, and a near-solar
metallicity in SY\,Mus.  The very low $^{12}$C/$^{13}$C isotopic ratios,
$\sim$6--10, derived for both objects and their N enrichment indicate that the
giants have experienced the first dredge-up.

%==================================================================

\section{Acknowledgments}

This study is supported by the Polish National Science Centre grant
DEC-2011/01/B/ST9/06145.  The observations were obtained at the
Gemini Observatory, which is operated by the Association of
Universities for Research in Astronomy, Inc., under a cooperative
agreement with the NSF on behalf of the Gemini partnership: the
National Science Foundation (United States), the National Research
Council (Canada), CONICYT (Chile), the Australian Research Council
(Australia), Minist\'{e}rio da Ci\^{e}ncia, Tecnologia e Inova\c{c}\~{a}o
(Brazil) and Ministerio de Ciencia, Tecnolog\'{i}a e Innovaci\'{o}n
Productiva (Argentina).  
K.H. acknowledges travel support from the NOAO Office of Science.
M.G. has been financed by the GEMINI-CONICYT Fund, allocated to the project 32110014.

%==================================================================

\appendix

\section[]{List of atomic and molecular lines}

\begin{table}
 \centering
  \caption{List of atomic lines selected for calculations
together with $gf$-values and excitation potentials.}\label{TLA}
  \begin{tabular}{@{}lccc@{}}
  \hline
Wavelength (air) & $EP$     & $\log{gf}$ & Ref. \\
$[\AA]$    & $[$eV$]$ &            &      \\
  \hline
\mbox{Sc\,{\sc i}}      &          &            & \\
$^{\ddagger}$22202.640  & 1.428    & -3.436     & \citet{Kup1999}\\%11
$^{\ddagger}$22207.128  & 1.428    & -1.836     & \citet{Kup1999}\\%11
22266.729  & 1.428    & -1.327     & \citet{Kup1999}\\%11
$^{\ddagger}$ 23603.945 & 1.428    & -2.180     & \citet{Kup1999}\\%11
\mbox{Ti\,{\sc i}}      &          &            & \\
15598.890  & 4.690    & -0.030     & \citet{MeBa1999}\\
15602.840  & 2.270    & -1.810     & \citet{MeBa1999}\\
22210.636  & 4.213    & -1.444     & \citet{Kup1999}\\%12
22211.228  & 1.734    & -1.770     & \citet{Kup1999}\\%12
22224.530  & 4.933    & -0.263     & \citet{Kup1999}\\%12
22232.838  & 1.739    & -1.658     & \citet{Kup1999}\\%12
22274.012  & 1.749    & -1.756     & \citet{Kup1999}\\%12
22282.973  & 4.690    & -0.586     & \citet{Kup1999}\\%12
22310.617  & 1.734    & -2.124     & \citet{Kup1999}\\%12
\mbox{Fe\,{\sc i}}      &          &            & \\
15593.740  & 5.030    & -1.980     & \citet{MeBa1999}\\
15598.870  & 6.240    & -0.920     & \citet{MeBa1999}\\
15604.220  & 6.240    &  0.280     & \citet{MeBa1999}\\
15613.630  & 6.350    & -0.290     & \citet{MeBa1999}\\
15621.160  & 6.200    & -0.990     & \citet{MeBa1999}\\
15621.650  & 5.540    & +0.170     & \citet{MeBa1999}\\
15629.370  & 5.950    & -1.670     & \citet{MeBa1999}\\
15629.630  & 4.560    & -3.130     & \citet{MeBa1999}\\
15631.110  & 3.640    & -3.980     & \citet{MeBa1999}\\
15631.950  & 5.350    & -0.150     & \citet{MeBa1999}\\
15645.010  & 6.310    & -0.540     & \citet{MeBa1999}\\
15647.410  & 6.330    & -1.090     & \citet{MeBa1999}\\
15648.520  & 5.430    & -0.800     & \citet{MeBa1999}\\
15652.870  & 6.250    & -0.190     & \citet{MeBa1999}\\
15656.640  & 5.870    & -1.900     & \citet{MeBa1999}\\
15662.010  & 5.830    & +0.000     & \citet{MeBa1999}\\
15662.320  & 6.330    & -1.020     & \citet{MeBa1999}\\
15665.240  & 5.980    & -0.600     & \citet{MeBa1999}\\
22239.040  & 5.385    & -3.121     & \citet{Kup1999}\\%2
22257.097  & 5.064    & -0.723     & \citet{Kup1999}\\%2
22260.185  & 5.086    & -0.952     & \citet{Kup1999}\\%2
$^{\ddagger}$23566.671  & 6.144    &  0.306     & \citet{Kup1999}\\%11
$^{\ddagger}$23572.267  & 5.874    & -0.886     & \citet{Kup1999}\\%11
\mbox{Ni\,{\sc i}}      &          &            & \\
15605.6800 & 5.300    & -0.590     & \citet{MeBa1999}\\
15605.7500 & 5.300    & -1.010     & \citet{MeBa1999}\\
15632.6200 & 5.310    & -0.130     & \citet{MeBa1999}\\
  \hline                             
\end{tabular}
\begin{list}{}{}
\item[$^{\ddagger}$] Not used for the chemical composition determination.
\end{list}                       
\end{table}                          

\begin{table}
 \centering  
  \caption{List of molecular lines selected for calculations
together with $gf$-values and excitation potentials.}\label{TLM}
\begin{tabular}{@{}lccc@{}}
%\begin{longtable}{@{}lccc@{}}
  \hline
Wavelength (air) & $EP$     & $\log{gf}$ & Ref. \\
$[\AA]$    & $[$eV$]$ &            &      \\
  \hline
$^{12}$C$^{14}$N     &          &            & \\
15594.607  & 0.9162   & -1.645     & \citet{Kur1999}\\%
15594.610  & 1.1079   & -1.240     & \citet{Kur1999}\\%
15594.816  & 1.1770   & -1.097     & \citet{Kur1999}\\%
15595.729  & 0.9218   & -2.106     & \citet{Kur1999}\\%
15597.719  & 1.3900   & -1.443     & \citet{Kur1999}\\%
15599.961  & 1.2391   & -1.807     & \citet{Kur1999}\\%
15605.217  & 1.2601   & -1.810     & \citet{Kur1999}\\%
15606.718  & 0.9095   & -2.116     & \citet{Kur1999}\\%
15607.198  & 1.5181   & -1.318     & \citet{Kur1999}\\%
15609.198  & 0.9361   & -1.479     & \citet{Kur1999}\\%
15613.554  & 1.0521   & -1.645     & \citet{Kur1999}\\%
15613.557  & 1.1770   & -1.090     & \citet{Kur1999}\\%
15615.246  & 1.0359   & -1.645     & \citet{Kur1999}\\%
15617.057  & 1.1080   & -1.231     & \citet{Kur1999}\\%
15617.646  & 1.3900   & -1.435     & \citet{Kur1999}\\%
15622.431  & 0.9162   & -1.637     & \citet{Kur1999}\\%
15624.806  & 1.5182   & -1.311     & \citet{Kur1999}\\%
15626.993  & 0.9345   & -2.088     & \citet{Kur1999}\\%
15629.867  & 1.2602   & -1.798     & \citet{Kur1999}\\%
15631.866  & 0.9362   & -1.473     & \citet{Kur1999}\\%
15636.015  & 0.9218   & -2.097     & \citet{Kur1999}\\%
15636.361  & 1.2816   & -1.801     & \citet{Kur1999}\\%
15636.584  & 1.1314   & -1.231     & \citet{Kur1999}\\%
15637.521  & 0.9357   & -1.633     & \citet{Kur1999}\\%
15640.865  & 1.4182   & -1.436     & \citet{Kur1999}\\%
15643.764  & 1.0688   & -1.632     & \citet{Kur1999}\\%
15643.965  & 1.0521   & -1.632     & \citet{Kur1999}\\%
15646.246  & 1.2059   & -1.090     & \citet{Kur1999}\\%
15658.785  & 1.1314   & -1.223     & \citet{Kur1999}\\%
15659.161  & 0.9476   & -2.070     & \citet{Kur1999}\\%
15660.014  & 1.5516   & -1.313     & \citet{Kur1999}\\%
15660.688  & 1.4182   & -1.428     & \citet{Kur1999}\\%
15660.705  & 1.2817   & -1.789     & \citet{Kur1999}\\%
15661.584  & 0.9612   & -1.471     & \citet{Kur1999}\\%
22194.477  & 1.2309   & -1.747     & \citet{Kur1999}\\%
22196.680  & 1.4573   & -1.875     & \citet{Kur1999}\\%
22198.914  & 1.4373   & -1.872     & \citet{Kur1999}\\%
22198.955  & 1.6007   & -1.825     & \citet{Kur1999}\\%
22216.477  & 1.2165   & -1.747     & \citet{Kur1999}\\% 
22220.176  & 1.1036   & -2.055     & \citet{Kur1999}\\% 
22228.924  & 1.1079   & -2.204     & \citet{Kur1999}\\% 
22228.994  & 1.3036   & -1.633     & \citet{Kur1999}\\% 
22236.898  & 1.1074   & -2.249     & \citet{Kur1999}\\% 
22239.814  & 1.6007   & -1.816     & \citet{Kur1999}\\% 
22243.514  & 1.2456   & -1.732     & \citet{Kur1999}\\% 
22248.297  & 1.4573   & -1.862     & \citet{Kur1999}\\% 
22248.438  & 1.4777   & -1.866     & \citet{Kur1999}\\% 
22262.354  & 1.2309   & -1.732     & \citet{Kur1999}\\% 
22273.969  & 1.6282   & -1.818     & \citet{Kur1999}\\% 
22276.160  & 1.3036   & -1.624     & \citet{Kur1999}\\% 
22277.008  & 1.1080   & -2.197     & \citet{Kur1999}\\% 
22280.246  & 1.1036   & -2.047     & \citet{Kur1999}\\% 
22285.527  & 1.1177   & -2.225     & \citet{Kur1999}\\% 
22294.160  & 1.2609   & -1.718     & \citet{Kur1999}\\% 
22295.695  & 1.1216   & -2.042     & \citet{Kur1999}\\% 
22296.930  & 1.3900   & -1.803     & \citet{Kur1999}\\% 
22299.426  & 1.4778   & -1.853     & \citet{Kur1999}\\% 
22300.430  & 1.1397   & -2.192     & \citet{Kur1999}\\% 
22301.910  & 1.4986   & -1.856     & \citet{Kur1999}\\% 
22302.410  & 1.3260   & -1.624     & \citet{Kur1999}\\% 
22309.961  & 1.2457   & -1.718     & \citet{Kur1999}\\% 
22311.473  & 1.2942   & -2.377     & \citet{Kur1999}\\% 
  \hline                             
\end{tabular}                        
%\end{longtable}                        
\end{table}

\begin{table}                           
 \centering                             
  \contcaption{}                        
\begin{tabular}{@{}lccc@{}}             
%\begin{longtable}{@{}lccc@{}}          
  \hline
Wavelength & $EP$     & $\log{gf}$ & Ref. \\
$[\AA]$    & $[$eV$]$ &            &      \\
  \hline
$^{12}$C$^{16}$O     &          &            & \\
15594.223  & 0.1313   & -7.7545    & \citet{Goo1994}\\
15595.946  & 0.5118   & -7.3273    & \citet{Goo1994}\\
15597.348  & 0.1204   & -7.7786    & \citet{Goo1994}\\
15599.257  & 0.5339   & -7.3121    & \citet{Goo1994}\\
15600.737  & 0.1100   & -7.8035    & \citet{Goo1994}\\
15602.842  & 0.5564   & -7.2971    & \citet{Goo1994}\\
15604.392  & 0.1000   & -7.8294    & \citet{Goo1994}\\
15606.702  & 0.5794   & -7.2822    & \citet{Goo1994}\\
15608.312  & 0.0905   & -7.8564    & \citet{Goo1994}\\
15612.497  & 0.0814   & -7.8841    & \citet{Goo1994}\\
15615.250  & 0.6268   & -7.2530    & \citet{Goo1994}\\
15616.948  & 0.0729   & -7.9133    & \citet{Goo1994}\\
15619.940  & 0.6512   & -7.2385    & \citet{Goo1994}\\
15621.663  & 0.0648   & -7.9439    & \citet{Goo1994}\\
15624.908  & 0.6760   & -7.2242    & \citet{Goo1994}\\
15626.644  & 0.0572   & -7.9755    & \citet{Goo1994}\\
15630.155  & 0.7012   & -7.2101    & \citet{Goo1994}\\
15631.891  & 0.0500   & -8.0092    & \citet{Goo1994}\\
15635.682  & 0.7269   & -7.1961    & \citet{Goo1994}\\
15637.404  & 0.0434   & -8.0446    & \citet{Goo1994}\\
15641.490  & 0.7531   & -7.1822    & \citet{Goo1994}\\
15643.182  & 0.0372   & -8.0824    & \citet{Goo1994}\\
15647.580  & 0.7797   & -7.1685    & \citet{Goo1994}\\
15649.227  & 0.0314   & -8.1226    & \citet{Goo1994}\\
15653.953  & 0.8068   & -7.1548    & \citet{Goo1994}\\
15655.537  & 0.0262   & -8.1659    & \citet{Goo1994}\\
15660.610  & 0.8343   & -7.1412    & \citet{Goo1994}\\
15662.115  & 0.0214   & -8.2128    & \citet{Goo1994}\\
23554.786  & 0.0048   & -6.4587    & \citet{Goo1994}\\
23556.234  & 1.4594   & -4.2830    & \citet{Goo1994}\\
23558.269  & 0.8385   & -4.5974    & \citet{Goo1994}\\
23562.231  & 1.4886   & -4.2734    & \citet{Goo1994}\\
23564.307  & 0.8218   & -4.6121    & \citet{Goo1994}\\
23564.951  & 2.1495   & -4.3503    & \citet{Goo1994}\\
23568.674  & 1.5182   & -4.2638    & \citet{Goo1994}\\
23569.863  & 2.5598   & -4.7140    & \citet{Goo1994}\\
23570.723  & 0.2870   & -5.5423    & \citet{Goo1994}\\
23570.766  & 0.8056   & -4.6271    & \citet{Goo1994}\\
23575.563  & 1.5482   & -4.2545    & \citet{Goo1994}\\
23577.647  & 0.7898   & -4.6424    & \citet{Goo1994}\\
23577.693  & 0.0071   & -6.3642    & \citet{Goo1994}\\
23582.901  & 1.5786   & -4.2451    & \citet{Goo1994}\\
23582.995  & 2.1902   & -4.3426    & \citet{Goo1994}\\
23584.949  & 0.7744   & -4.6582    & \citet{Goo1994}\\
23587.982  & 0.2827   & -5.5907    & \citet{Goo1994}\\
23590.689  & 1.6095   & -4.2359    & \citet{Goo1994}\\
23592.672  & 0.7596   & -4.6743    & \citet{Goo1994}\\
23594.794  & 2.6067   & -4.7071    & \citet{Goo1994}\\
23598.928  & 1.6408   & -4.2268    & \citet{Goo1994}\\
23600.817  & 0.7451   & -4.6908    & \citet{Goo1994}\\
23601.032  & 0.0100   & -6.2876    & \citet{Goo1994}\\
23601.535  & 2.2314   & -4.3350    & \citet{Goo1994}\\
23605.665  & 0.2789   & -5.6445    & \citet{Goo1994}\\
23607.621  & 1.6725   & -4.2177    & \citet{Goo1994}\\
23609.383  & 0.7312   & -4.7077    & \citet{Goo1994}\\
23616.77   & 1.7047   & -4.2088    & \citet{Goo1994}\\
23618.369  & 0.7177   & -4.7251    & \citet{Goo1994}\\
23620.259  & 2.6541   & -4.7001    & \citet{Goo1994}\\
23620.571  & 2.2729   & -4.3273    & \citet{Goo1994}\\
23623.771  & 0.2756   & -5.7051    & \citet{Goo1994}\\
23624.802  & 0.0133   & -6.2232    & \citet{Goo1994}\\
23626.375  & 1.7373   & -4.2000    & \citet{Goo1994}\\
23627.778  & 0.7046   & -4.7430    & \citet{Goo1994}\\
  \hline                             
\end{tabular}                        
%\end{longtable}                        
\end{table}

\begin{table}
 \centering  
  \contcaption{}
\begin{tabular}{@{}lccc@{}}
%\begin{longtable}{@{}lccc@{}}
  \hline
Wavelength (air) & $EP$     & $\log{gf}$ & Ref. \\
$[\AA]$    & $[$eV$]$ &            &      \\
  \hline
23636.44   & 1.7703   & -4.1912    & \citet{Goo1994}\\
23637.608  & 0.6921   & -4.7617    & \citet{Goo1994}\\
23640.108  & 2.3148   & -4.3198    & \citet{Goo1994}\\
23642.301  & 0.2728   & -5.7747    & \citet{Goo1994}\\
23646.262  & 2.7018   & -4.6931    & \citet{Goo1994}\\
23646.967  & 1.8037   & -4.1824    & \citet{Goo1994}\\
23647.859  & 0.6799   & -4.7807    & \citet{Goo1994}\\
23649.006  & 0.0172   & -6.1676    & \citet{Goo1994}\\
$^{12}$C$^{17}$O     &          &            & \\
23555.604  & 0.5678   & -4.9176    & \citet{Goo1994}\\
23558.862  & 0.0255   & -5.9974    & \citet{Goo1994}\\
23561.873  & 0.5513   & -4.9326    & \citet{Goo1994}\\
23568.546  & 0.5353   & -4.9473    & \citet{Goo1994}\\
23575.626  & 0.5198   & -4.9630    & \citet{Goo1994}\\
23583.110  & 0.5047   & -4.9788    & \citet{Goo1994}\\
23591.000  & 0.4901   & -4.9948    & \citet{Goo1994}\\
23599.294  & 0.4759   & -5.0114    & \citet{Goo1994}\\
23607.995  & 0.4621   & -5.0284    & \citet{Goo1994}\\
23617.100  & 0.4489   & -5.0459    & \citet{Goo1994}\\
23626.613  & 0.4360   & -5.0639    & \citet{Goo1994}\\
23636.530  & 0.4237   & -5.0825    & \citet{Goo1994}\\
23646.853  & 0.4117   & -5.1016    & \citet{Goo1994}\\
$^{12}$C$^{18}$O     &          &            & \\
23556.518  & 0.2538   & -5.4594    & \citet{Goo1994}\\
23564.155  & 0.2389   & -5.4752    & \citet{Goo1994}\\
23572.184  & 0.2245   & -5.4913    & \citet{Goo1994}\\
23580.602  & 0.2105   & -5.5079    & \citet{Goo1994}\\
23589.412  & 0.1970   & -5.5251    & \citet{Goo1994}\\
23598.614  & 0.1839   & -5.5426    & \citet{Goo1994}\\
23608.206  & 0.1712   & -5.5607    & \citet{Goo1994}\\
23618.19   & 0.1590   & -5.5792    & \citet{Goo1994}\\
23628.565  & 0.1473   & -5.5984    & \citet{Goo1994}\\
23639.333  & 0.1360   & -5.6182    & \citet{Goo1994}\\
$^{13}$C$^{16}$O     &          &            & \\
23560.596  & 1.3099   & -4.9751    & \citet{Goo1994}\\
23563.521  & 0.1719   & -5.5580    & \citet{Goo1994}\\
23570.753  & 1.3437   & -4.9666    & \citet{Goo1994}\\
23573.485  & 0.1596   & -5.5766    & \citet{Goo1994}\\
23581.341  & 1.3779   & -4.9582    & \citet{Goo1994}\\
23583.842  & 0.1478   & -5.5957    & \citet{Goo1994}\\
23592.363  & 1.4126   & -4.9496    & \citet{Goo1994}\\
23594.59   & 0.1365   & -5.6156    & \citet{Goo1994}\\
23603.818  & 1.4476   & -4.9412    & \citet{Goo1994}\\
23605.733  & 0.1256   & -5.6360    & \citet{Goo1994}\\
23615.711  & 1.4831   & -4.9329    & \citet{Goo1994}\\
23617.268  & 0.1151   & -5.6574    & \citet{Goo1994}\\
23628.043  & 1.5189   & -4.9248    & \citet{Goo1994}\\
23629.197  & 0.1051   & -5.6794    & \citet{Goo1994}\\
23640.814  & 1.5552   & -4.9165    & \citet{Goo1994}\\
23641.522  & 0.0956   & -5.7025    & \citet{Goo1994}\\
OH         &          &            & \\
15593.563  & 0.8740   & -5.358     & \citet{Kur1999}\\%
15608.357  & 0.4942   & -7.209     & \citet{Kur1999}\\%
15609.683  & 0.4942   & -7.209     & \citet{Kur1999}\\%
15621.766  & 0.8374   & -6.734     & \citet{Kur1999}\\%
15624.434  & 0.8415   & -7.006     & \citet{Kur1999}\\%
15624.660  & 0.1336   & -8.233     & \citet{Kur1999}\\%
15626.704  & 0.5413   & -5.198     & \citet{Kur1999}\\%
15627.290  & 0.8871   & -5.435     & \citet{Kur1999}\\%
15627.293  & 0.8871   & -5.435     & \citet{Kur1999}\\%
15627.413  & 0.5413   & -5.198     & \citet{Kur1999}\\%
15628.902  & 0.1337   & -8.233     & \citet{Kur1999}\\%
15636.235  & 0.8876   & -7.202     & \citet{Kur1999}\\%
  \hline                             
\end{tabular}                           
%\end{longtable}                        
\end{table}

\begin{table}
 \centering  
  \contcaption{}
\begin{tabular}{@{}lccc@{}}
%\begin{longtable}{@{}lccc@{}}
  \hline
Wavelength (air) & $EP$     & $\log{gf}$ & Ref. \\
$[\AA]$    & $[$eV$]$ &            &      \\
  \hline
15636.596  & 0.8876   & -7.202     & \citet{Kur1999}\\%
15643.302  & 0.8420   & -7.007     & \citet{Kur1999}\\%
15650.557  & 0.8643   & -5.587     & \citet{Kur1999}\\%
15650.797  & 0.8643   & -5.587     & \citet{Kur1999}\\%
15651.896  & 0.5341   & -5.132     & \citet{Kur1999}\\%
15653.480  & 0.5343   & -5.132     & \citet{Kur1999}\\%
15654.116  & 0.8383   & -6.734     & \citet{Kur1999}\\%
15655.053  & 0.3041   & -7.713     & \citet{Kur1999}\\%
15658.127  & 0.3038   & -7.713     & \citet{Kur1999}\\%
  \hline                             
\end{tabular}                        
%\end{longtable}                      
\end{table}                          
                                     
%content                             
                                     
%
%\newpage
%

%\subsection{Subsection title}

\bsp

\label{lastpage}


\begin{thebibliography}{99}

\bibitem[\protect\citeauthoryear{Aboussaid et al.}{2006}]{aboussaid_et_al_2006} Aboussaid, A., Carleer, M., Hurtmans, D.,
Bi\'emont, E., \& Godefroid, M.R. 2006, Physica Scripta, 53, 28 

\bibitem[\protect\citeauthoryear{Asplund et al.}{2009}]{Asp2009} Asplund, M., Grevesse, N., Sauval A., \& Scott, P., 2009, 
ARAA 47, 481 %

\bibitem[\protect\citeauthoryear{Belczy\'nski et al.}{2000}]{Bel2000} Belczy\'nski, K., Miko{\l}ajewska, J., Munari, U., et al., 2000, 
A\&AS, 146, 407 %

\bibitem[\protect\citeauthoryear{Brandt}{1998}]{Bra1998} Brandt, S., Data Analysis, Statistical and Computational Methods, 1998,
Polish edition (Polish Scientific Publishers PWN) %

\bibitem[\protect\citeauthoryear{Carlberg et al.}{2011}]{Car2011} Carlberg, J. K., Majewski, S. R., Patterson, R. J., et al., 2011,
ApJ, 732, 39 %

\bibitem[\protect\citeauthoryear{Cunha \& Smith}{2006}]{CuSm2006} Cunha, K., \& Smith, V. V., 2006,
ApJ, 651, 491 %

\bibitem[\protect\citeauthoryear{Davidsen et al.}{1977}]{Dav1977} Davidsen, A., Malina, R., Bowyer, S., 1977,
ApJ, 211, 866 %

\bibitem[\protect\citeauthoryear{Dumm et al.}{1998}]{Dum1999} Dumm, T., Schmutz, W., Schild, H., Nussbaumer, H., 1999, 
A\&A, 349, 169 %

\bibitem[\protect\citeauthoryear{Fekel et al.}{2003}]{Fek2003} Fekel, F. C., Hinkle, H. K., Joyce, R. R., et al., 2003, 
ASPC, 303, 113 %

\bibitem[\protect\citeauthoryear{Garcia et al.}{1983}]{Gar1983} Garcia, M., Baliunas, S. L., Elvis, M., et al., 1983, 
ApJ, 267, 291 %

\bibitem[\protect\citeauthoryear{Goorvitch}{1994}]{Goo1994} Goorvitch, D., 1994,
ApJS, 95, 535 %

\bibitem[\protect\citeauthoryear{Gustafsson et al.}{2008}]{Gus2008} Gustafsson, B., Edvardsson, B., Eriksson, K., et al., 2008,
A\&A, 486, 951 %

\bibitem[\protect\citeauthoryear{Hinkle et al.}{2006}]{Hin2006} Hinkle, K. H., Fekel, F. C., Joyce, R. R., et al., 2006,
ApJ, 641, 479 %

\bibitem[\protect\citeauthoryear{Hog et al.}{2003}]{hog2000} Hog, E., Fabricius, C., Makarov, et al., 2000, A\&A, 355, L27%

\bibitem[\protect\citeauthoryear{Jorissen}{2003}]{Jor2003} Jorissen, A., 2003,
ASPC, 303, 25 %

\bibitem[\protect\citeauthoryear{Joyce}{1992}]{Joy1992} Joyce, R. 1992, in ASP Conf. Ser. 23, Astronomical CCD Observing and
Reduction Techniques, ed. S. Howell (San Francisco: ASP), 258 %

\bibitem[\protect\citeauthoryear{Kallrath \& Milone}{1999}]{KaMi1999} Kallrath, J., \& Milone, E. F., Eclipsing binary stars – modeling and analysis, 1999, (New York: Springer) %

\bibitem[\protect\citeauthoryear{Keller et al.}{2001}]{Kel2001} Keller, L. D., Pilachowski, C. A., Sneden, C., 2001, 
AJ, 122, 2554 %

\bibitem[\protect\citeauthoryear{Kenyon \& Miko{\l}ajewska}{1995}]{KeMi1995} Kenyon, S. J., \& Miko{\l}ajewska, J., 1995, 
AJ, 110, 391 %

\bibitem[\protect\citeauthoryear{Kucinskas et al.}{2005}]{Kuc2005} Kucinskas, A., Hauschildt, P. H., Ludwig, H.-G., et al., 2005,
A\&A, 442, 281 %

\bibitem[\protect\citeauthoryear{Kupka et al.}{1999}]{Kup1999} Kupka, F., Piskunov, N., Ryabchikova, T., et al., 1999,
A\&AS, 138, 119 %

\bibitem[\protect\citeauthoryear{Kurucz}{1999}]{Kur1999} Kurucz, R. L., 1999, {\it{http://kurucz.harvard.edu}}

\bibitem[\protect\citeauthoryear{M\'elendez \& Barbuy}{1999}]{MeBa1999} M\'elendez, J., \& Barbuy, B., 1999,
ApJS, 124, 527 %

\bibitem[\protect\citeauthoryear{Miko{\l}ajewska}{2003}]{Mik2003} Miko{\l}ajewska, J. 2003, in "Symbiotic Stars Probing Stellar Evolution",
ASP Conference Proceedings, eds. R. L. M. Corradi, J. Miko{\l}ajewska and T. Mahoney, 303, p.9. %

\bibitem[\protect\citeauthoryear{M{\"u}rset \& Schmid}{1999}]{Mue1999} M{\"u}rset, U., \& Schmid, H. M., 1999,
A\&AS, 137, 473 %

\bibitem[\protect\citeauthoryear{Nussbaumer et al.}{1988}]{Nus1988} Nussbaumer, H., Schild, H., Schmid, H. M., Vogel, M., 1988,
A\&A, 198, 179 %

\bibitem[\protect\citeauthoryear{Otulakowska-Hypka et al.}{2013}]{Otu2013} Otulakowska-Hypka, M., Miko{\l}ajewska, J., Whitelock, P., 2013, in {\it{Stella Novae: Future and Past Decades}},
ASPC, in press %

\bibitem[\protect\citeauthoryear{Pehlivan}{2012}]{Peh2012} Pehlivan Asli, 2012, "Laboratory Spectroscopy of Neutral Scandium, Sc\'I, for Astrophysical Application"
MSc thesis at Lund Observatory%

\bibitem[\protect\citeauthoryear{Pereira et al.}{1998}]{Per1998} Pereira, C. B., Smith, V. V., Cunha, K., 1998,
AJ, 116, 1977 %

\bibitem[\protect\citeauthoryear{Pereira et al.}{2005}]{Per2005} Pereira, C. B., Smith, V. V., Cunha, K., 2005,
A\&A, 429, 993 %

\bibitem[\protect\citeauthoryear{Richichi et al.}{1999}]{Ric1999} Richichi, A., Fabbroni, L., Ragland, S., Scholz, M., 1999,
A\&A, 344, 511 %

\bibitem[\protect\citeauthoryear{Rutkowski et al.}{2007}]{Rut2007} Rutkowski, A., Miko{\l}ajewska, J., \& Whitelock, P., 2007,
Baltic Astronomy, 16, 49 %

\bibitem[\protect\citeauthoryear{Schild et al.}{1996}]{Sch1996} Schild, H., Murset, U., \& Schmutz, W., 1996, 
A\&A, 306, 477 %

\bibitem[\protect\citeauthoryear{Schild et al.}{1992}]{Sch1992} Schild, H., Boyle, S. J., Schmid, H. M., 1992, 
MNRAS, 258, 95 %

\bibitem[\protect\citeauthoryear{Schmidt \& Miko{\l}ajewska}{2003}]{Sch2003} Schmidt, M. R., \& Miko{\l}ajewska, J., 2003,
ASPC, 303, 163 %

\bibitem[\protect\citeauthoryear{Schmidt et al.}{2006}]{Sch2006} Schmidt, M. R., Zacs, L., Miko{\l}ajewska, J., \& Hinkle, K., 2006,
A\&A, 446, 603 %

\bibitem[\protect\citeauthoryear{Schmutz et al.}{1994}]{Sch1994} Schmutz, W., Schild, H., Murset, U., \& Schmid, H., 1994, 
ApJ, 556, 55 %

\bibitem[\protect\citeauthoryear{Smith \& Lambert}{1990}]{SmLa1986} Smith, V. V., \& Lambert, D., 1986,
ApJ, 311, 843 %

\bibitem[\protect\citeauthoryear{Smith \& Lambert}{1990}]{SmLa1990} Smith, V. V., \& Lambert, D., 1990,
ApJS, 72, 387 %

\bibitem[\protect\citeauthoryear{Smith et al.}{1996}]{Smi1996} Smith, V. V., Cunha, K., Jorissen, A., Boffin, H. M. J., 1996,
A\&A, 315, 179 %

\bibitem[\protect\citeauthoryear{Smith et al.}{1997}]{Smi1997} Smith, V. V., Cunha, K., Jorissen, A., Boffin, H. M. J., 1997,
A\&A, 324, 97 %

\bibitem[\protect\citeauthoryear{Smith et al.}{2001}]{Smi2001} Smith, V. V., Pereira, C. B., Cunha, K., 2001,
ApJ, 556, 55 %

\bibitem[\protect\citeauthoryear{Smith et al.}{2002}]{Smi2002} Smith, V. V., Hinkle, K. H., Cunha, K., 2002,
AJ, 124, 3241 %

\bibitem[\protect\citeauthoryear{Tifft \& Greenstein}{1958}]{Tif1958} Tifft, W. G., \& Greenstein, J. L., 1958,
ApJ, 127, 160 %

\bibitem[\protect\citeauthoryear{Wallerstein et al.}{2008}]{Wal2008} Wallerstein, G., Harrison, T., Munari, U., Vanture, A., 2008,
PASP, 120, 492 %

\bibitem[\protect\citeauthoryear{Wiese \& Fuhr}{1975}]{WiFu1975} Wiese, W. L., \& Fuhr, J. R., 1975,
J. Phys. Chem. Ref. Data, 4, 263 %

%VALD - cytacje:
%If VALD has been used in your research work, you agree to reference it in
%your publications. Up to now VALD and its development have been described in
%the following papers:
%    Kupka F., Piskunov N.E., Ryabchikova T.A., Stempels H.C., Weiss W.W.,
%A&AS 138, 119-133 (1999), (VALD-2)
%    Ryabchikova T.A. Piskunov N.E., Stempels H.C., Kupka F., Weiss W.W.
%Proc. of the 6th International Colloquium on Atomic Spectra and Oscillator
%Strengths, Victoria BC, Canada, 1998, Physica Scripta T83, 162-173 (1999),
%(VALD-2)
%    Piskunov N.E., Kupka F., Ryabchikova T.A., Weiss W.W., Jeffery C.S.,
%A&AS 112, 525 (1995) (VALD-1) 

%You also agree to reference to original source(s) of the line data that are
%important for your research. A complete list of references is available from
%the bibliography section of the VALD manual. A short list of references is
%compiled and sent to you with every reply from VALD. 

\end{thebibliography}
\end{document}